\pdfoutput=1
\documentclass[journal]{IEEEtran}
\usepackage{amsmath,amsfonts}
\usepackage{graphicx}
\DeclareGraphicsExtensions{.pdf,.png,.jpg}
\usepackage{booktabs}
\usepackage{algorithm}
\usepackage{algpseudocode}
\usepackage{array}
\usepackage{textcomp}
\usepackage{stfloats}
\usepackage{url}
\usepackage{verbatim}
\usepackage{braket}

\newcommand{\ketbra}[2]{|{#1}\rangle\langle{#2}|}
\newcommand\thefont{\expandafter\string\the\font}

\usepackage{color}
\usepackage[colorinlistoftodos]{todonotes}
\usepackage{soul,xcolor}
\usepackage[unicode]{hyperref}
\begin{document}

\title{Quantum-Inspired Optimization through Qudit-Based Imaginary Time Evolution}

\author{

Erik M. Åsgrim$^{2}$,
Ahsan Javed Awan$^{1}$

\vspace{0.2cm} \normalsize $^1$Ericsson Research,
$^2$KTH Royal Institute of Technology

erima@kth.se, 
ahsan.javed.awan@ericsson.com 
}

\maketitle

\begin{abstract}
Imaginary-time evolution has been shown to be a promising framework for tackling combinatorial optimization problems on quantum hardware. In this work, we propose a classical quantum-inspired strategy for solving combinatorial optimization problems with integer-valued decision variables by encoding decision variables into multi-level quantum states known as \textit{qudits}. This method results in a reduced number of decision variables compared to binary formulations while inherently incorporating single-association constraints. Efficient classical simulation is enabled by constraining the system to remain in a product state throughout optimization. The qudit states are optimized by applying a sequence of unitary operators that iteratively approximate the dynamics of imaginary time evolution. Unlike previous studies, we propose a gradient-based method of adaptively choosing the Hermitian operators used to generate the state evolution at each optimization step, as a means to improve the convergence properties of the algorithm. The proposed algorithm demonstrates promising results on 
Min-$d$-Cut problem with constraints, outperforming Gurobi on penalized constraint formulation, particularly for larger values of $d$.
\end{abstract}

\begin{IEEEkeywords}
Quantum inspired, combinatorial optimization, imaginary time evolution.
\end{IEEEkeywords}

\section{Introduction}
\IEEEPARstart{T}{he} development of hybrid quantum-classical algorithms designed to solve combinatorial optimization problems has seen significant advancements during the last decade. Many promising methods have been proposed, such as the \textit{quantum approximate optimization algorithm} (QAOA) \cite{farhi_quantum_2014}, \textit{variational quantum eigensolver} (VQE) \cite{peruzzo_variational_2014}, and various implementations of \textit{quantum imaginary time evolution} (QITE) \cite{motta_determining_2020,morris_performant_2024}. In a recent review article, Abbas \textit{et al.} \cite{abbas_challenges_2024} emphasize that assessing the potential of these algorithms requires moving beyond theoretical complexity arguments to focus on their practical performance in realistic problem settings. They highlight the importance of identifying which algorithmic features and problem structures might enable genuine improvements over classical heuristics.

Whether a given quantum approach can in fact provide a practical advantage depends critically on whether the states and dynamics it uses are genuinely hard to simulate classically. As numerous quantum optimization algorithms have been effectively simulated through classical approaches \cite{medvidovic_classical_2021,xu_mps-vqe_2024,li_variational_2024}, often employing tensor network techniques, the threshold for achieving quantum advantage has been raised. At the same time, these developments have spurred a complementary line of research into \textit{quantum-inspired optimization}, where techniques originally devised for quantum hardware are adapted into powerful new classical heuristics. Consequently, quantum-inspired methods are important both as a benchmark to assess quantum algorithms and as a source of new approaches for solving classical optimization problems.

Recent work has shown that QITE is not only a promising quantum optimization strategy but also a basis for developing quantum-inspired classical algorithms. In its original formulation, QITE simulates imaginary-time evolution using a sequence of unitary gates \cite{motta_determining_2020}. Each gate advances the state over a short imaginary-time interval, and longer evolutions are obtained by iterating this procedure. Although this typically results in deep circuits and high entanglement, which are not tractable for classical simulation, a recent adaptation demonstrated that restricting the evolution to product states enables efficient, polynomial-time classical simulation \cite{alam_solving_2023}. Surprisingly, a follow-up study found that allowing entangling two-qubit gates offers only small performance improvements \cite{bauer_combinatorial_2023}, suggesting that simplified, classically simulable versions of QITE can still achieve competitive optimization performance.

\begin{figure}
    \centering
    \includegraphics[width=\linewidth,
                     trim=.2cm 0 0 0,clip]{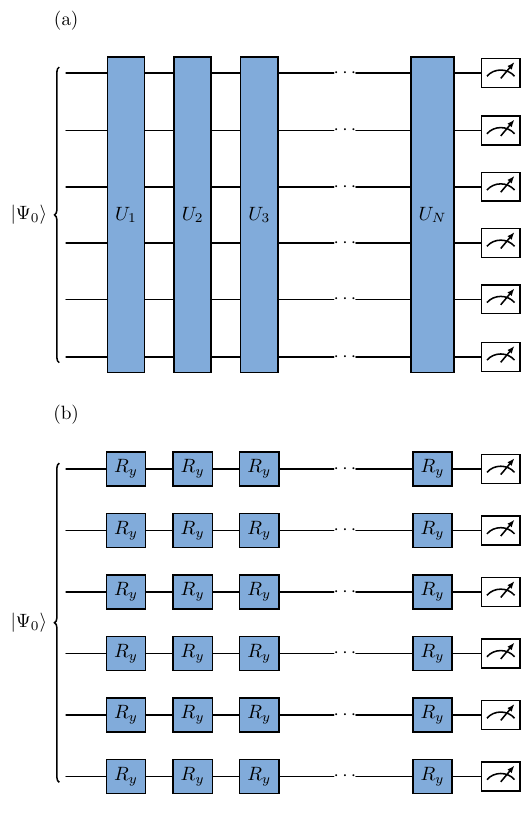}
    \caption{\textbf{(a)} Imaginary time evolution, as proposed by Motta et al. aims to approximate the imaginary time evolution by applying a sequence of unitary gates \protect\cite{motta_determining_2020}. Each unitary gate approximately propagates the quantum state for some finite imaginary time interval $\Delta\tau$. \textbf{(b)} Alam et al. proposed an implementation of QITE relying solely on single-qubit $R_y$ gates \protect\cite{alam_solving_2023}. By operating on product states of qubits classical simulation of large problem instances is enabled, albeit with strongly limited expressibility of the quantum state.}

    \label{fig-QITE_circuit}
\end{figure}
In this study, we build upon the idea of simulating QITE with product states, extending simulations from qubit two-level systems to multi-level quantum states known as \textit{qudits}. This generalization allows combinatorial optimization problems with integer-valued decision variables to be encoded directly into a set of qudit states. By doing so, we reduce the number of required decision variables compared to binary qubit-based encodings and eliminate the need for explicitly enforcing single-association constraints, thereby simplifying the optimization model. We approximate the imaginary-time evolution using unitary single-qudit operators, ensuring the state remains a product state throughout the process. In addition, rather than relying on a fixed set of Hermitian generators for the evolution, we introduce a gradient-based strategy that adaptively selects the Hermitian operator for each qudit at every optimization step in order to improve convergence.

To evaluate our method, we solve the Min-$d$-Cut problem with capacity constraints, setting an upper limit on the number of vertices in each partition. We benchmark the qudit-based QITE algorithm against Gurobi operating with binary decision variables. The experiments are performed on 10-regular graphs with $N=50,100,150$ vertices and $d=3,5,7$ partitions. The numerical results indicate promising performance, particularly for configurations with $d=7$ partitions, where our method achieves superior solutions on average compared to Gurobi when both algorithms utilize the same penalty-based constraint encoding. In summary, the main contributions of this work are as follows:
\begin{itemize}
    \item We extend QITE simulations from qubits to multi-level qudits, enabling direct encoding of integer-valued decision variables.
    \item We introduce a gradient-based method that adaptively chooses the Hermitian generators for single-qudit gates during optimization.
    \item We demonstrate that qudit-based encoding reduces the number of required decision variables and removes the need for explicit single-association constraints.
    \item We benchmark the qudit-based QITE algorithm on the Min-$d$-Cut problem with capacity constraints, showing promising performance and, in some cases, outperforming Gurobi.
\end{itemize}

\section{Background}
\subsection{Quantum imaginary time evolution (QITE)}
QITE describes the quantum state dynamics that are acquired by replacing the real time variable $t$ in the time-dependent Schrödinger equation by an imaginary time variable $\tau = it$. Throughout imaginary time, the state evolution generated by a Hamiltonian $H$ becomes
\begin{equation}
    \ket{\Psi(\tau)}= \frac{e^{-\tau H}\ket{\Psi_0}}{\sqrt{\braket{\Psi_0|e^{-2\tau H}|\Psi_0}}}.
    \label{eq-QITE_exact_evolution}
\end{equation} 

Where $\ket{\Psi_0}$ denotes the initial state and the denominator ensures that the state remains normalized. Throughout state evolution, the state $\ket{\Psi(\tau)}$ will effectively cool down to the ground state, given that the initial state $\ket{\Psi_0}$ has finite overlap with the ground state $|\braket{\Psi_0|\Psi_{\text{gs}}}|\neq 0$. This property makes QITE suitable for solving combinatorial optimization problems in which the ground state encodes the optimum solution. A well-established method of estimating the imaginary time evolution in \eqref{eq-QITE_exact_evolution} is to find a unitary operator $U$ that approximates the imaginary time evolution over an interval $\Delta \tau$, such that $\ket{\Psi(\tau + \Delta\tau)}\approx U(\Delta\tau)\ket{\Psi(\tau)}$ \cite{motta_determining_2020}. We let $U$ be generated by the Hermitian operator $G$ such that $U(\Delta \tau) =\exp(-i\Delta\tau G)$. Furthermore, we let the generator $G$ be given by a linear combination of a set of Hermitian operators $A=\{G_i\}$
\begin{equation}
    G = \sum_i a_i G_i
    \label{eq-G_linear_combination}
\end{equation}
for which the expansion coefficients $a_i$ are all real-valued. In this context, we refer to the set $A$ as the \textit{ansatz}, and it is specifically called a \textit{linear ansatz} when it consists solely of Hermitian operators generating non-entangling single-qubit gates. A unitary operator $U$ that approximates QITE over the interval $\Delta\tau$ can be found by minimizing the norm
\begin{equation}
    \delta = \left|\left|(1-i\Delta\tau G)\ket{\Psi(\tau)}-(1-\Delta\tau H)\ket{\Psi(\tau)}\right|\right|
    \label{eq-distance}
\end{equation}
under real variations of the expansion coefficients in \eqref{eq-G_linear_combination}. In previous work \cite{motta_determining_2020}, it has been shown that the expansion coefficients minimizing the norm $\delta$ in \eqref{eq-distance} satisfy the linear system of equations
\begin{equation}
    \sum_{i}\braket{\Psi(\tau)|G_i G_j|\Psi(\tau)}a_i = -\frac{i}{2}\braket{\Psi(\tau)|[H, G_j]|\Psi(\tau)}.
    \label{eq-linear_system_of_eqs}
\end{equation}
After the expansion coefficients have been acquired, the state can be propagated by applying the time evolution operator $U(\Delta\tau) = \exp(-i\Delta\tau\sum_i a_i G_i)$. It is possible to approximate QITE over longer imaginary time intervals than $\Delta \tau$, by recursively repeating the state update
\begin{equation}
    \ket{\Psi_{s+1}} = \exp\left(-i\Delta\tau\sum_i a_i[s] G_i\right)\ket{\Psi_s}.
    \label{eq-recursive_state_update}
\end{equation}
In \eqref{eq-recursive_state_update}, the coefficients $a_i[s]$ are determined by solving \eqref{eq-linear_system_of_eqs} with expectation values calculated relative to $\ket{\Psi_s}$, where $\ket{\Psi_s}$ represents the state at the $s^{th}$ optimization step.

\subsection{Related work}
Recently, an implementation of unitary-based QITE operating on product states of qubits was proposed, thus enabling efficient polynomial time classical simulation \cite{alam_solving_2023}. In the proposed method, the quantum state is initialized as a product state and the unitary time evolution is constrained to act as single-qudit $R_y$ rotation gates, as depicted in Fig. \ref{fig-QITE_circuit}(b). As the $R_y$ operators in the subsequent layers all commute, the problem of growing circuit depth is also mitigated. Although representing the quantum state trajectory during imaginary time evolution as a product state is naturally a crude model, a subsequent study demonstrated only limited improvements to performance when instead considering an ansatz that also allows for entangling two-qubit gates \cite{bauer_combinatorial_2023}.

Formally, the authors suggest using an ansatz of Pauli-Y operators $A = \{Y_i\}_{i=0}^{N-1}$, where $Y_i$ acts as the Pauli-Y operator on the $i$:th qubit and as the identity operator on the remaining qubits. Using this ansatz, the full Hermitian generator becomes
\begin{equation}
    G = \sum_{i = 0}^{N-1}a_i Y_i.
    \label{eq-qubit_linear_ansatz}
\end{equation}
Operating with the linear ansatz in \eqref{eq-qubit_linear_ansatz}, the unitary $U$ will become a product of single-qubit $R_y$ gates
\begin{equation}
    U(\Delta\tau) = \exp\left(-i\Delta\tau\sum_i a_i Y_i\right)=\bigotimes_i R^i_y\left(\theta_i\right)
    \label{eq-Ry_prod_unitary}
\end{equation}
where we let $R_y^i$ denote a rotation gate on the $i$:th qubit. In \eqref{eq-Ry_prod_unitary}, the rotation angles are related to the expansion coefficients as $\theta_i = 2a_i\Delta\tau$. Given that the initial state $\ket{\Psi_0}$ is in a product state, the state will remain in a product state throughout imaginary time evolution, allowing for efficient classical simulation. 

\section{Method}
In the following section, we propose a formulation of the QITE algorithm operating on product states of qudits, as a means to solve combinatorial optimization problems with integer-valued decision variables. In addition, we propose a protocol for adaptively choosing the linear ansatz at each optimization step. As a testbed for our proposed solution, we consider the Min-$d$-Cut problem with capacity constraints.

\subsection{Variable and cost encoding}
We consider an optimization problem consisting of the classical integer-valued decision variables $\{x_i\}_{i=0}^{N-1}$, where each variable can assume the values $x_i \in\{0,1,...,d-1\}$. We encode the \textit{i}:th variable in the normalized \textit{d}-level qudit state
\begin{equation}
    \ket{\psi_i} = \sum_{k=0}^{d-1} a_{i,k} \ket{k}
    \label{eq-qudit_state}
\end{equation}
using the orthonormal computational basis $\mathcal{B}_d = \{\ket{k}\}_{k=0}^{d-1}$. The full system of \textit{N} qudits is initially encoded in the product state
\begin{equation}
    \ket{\Psi_0} = \ket{0}\bigotimes_{i = 1}^{N-1}\frac{1}{\sqrt{d}}\sum_{k=0}^{d-1}\ket{k}.
    \label{eq-initial_state_qudit}
\end{equation} 
As the cost of Min-$d$-Cut is invariant under permutations of the state labels, the initial state \eqref{eq-initial_state_qudit} is guaranteed to have a finite overlap with the ground state subspace, despite assigning one qudit a definitive computational basis state. The classical cost function $C(\textbf{x})$ of the Min-$d$-Cut problem on a graph with edges $\langle i, j\rangle$ and edge weights $W_{i,j}$ is expressed as
\begin{equation}
    C(\textbf{x}) = \sum_{\langle i,j \rangle}W_{i,j}\left(1-\sum_{k=0}^{d-1} \delta_{x_i,k}\delta_{x_j, k}\right)
    \label{eq-min_d_cut_cost_integer}
\end{equation}
where $\delta$ denotes the Kronecker delta. The classical cost can conveniently be converted to a quantum Hamiltonian by replacing Kronecker deltas with computational basis state projectors $P_{i,k} = |k\rangle_i\langle k|_i$. Using this notation, $P_{i,k}$ projects the $i$:th qudit onto the $k$:th computational basis state, such that the expectation value $\braket{\Psi|P_{i,k}|\Psi}$ can be interpreted as the probability of qudit $i$ being in state $k$. Consequently, the expectation value $\braket{\Psi|\sum_{k=0}^{d-1} P_{i,k}P_{j,k}|\Psi}$ provides the probability that a given vertex pair $(i,j)$ is assigned to the same partition. Using this notation, the cost Hamiltonian can thus be expressed as
\begin{equation}
    H = \sum_{\langle i,j \rangle}W_{i,j}\left(1-\sum_{k=0}^{d-1} P_{i,k}P_{j, k}\right).
    \label{eq-max_d_cut_cost_hamiltonian}
\end{equation}
\subsection{Capacity constraints}
When solving the Min-$d$-Cut problem, we define a feasible solution as a set of integer-valued decision variables $\{x_i\}_{i=0}^{N-1}$ that satisfy the capacity constraint for every partition $k \in \{0, 1, ..., d-1\}$.

\begin{equation}
    \sum_{i=0}^{N-1} \delta_{x_i, k} \leq C_{\text{max}}
    \label{eq-capacity_constraint}
\end{equation}

 By including capacity constraints, we effectively impose the restriction that a maximum number of $C_{\text{max}}$ vertices can be assigned to any given partition. Given that the edge weights are all positive, capacity constraints are necessary to avoid a trivial solution to the Min-$d$-Cut problem, where assigning all vertices to the same partition generates the optimum solution. It is particularly interesting to investigate algorithm performance under capacity constraints, as many real-world optimization problems, such as knapsack-style problems \cite{bozejko_optimal_2024} and vehicle routing problems \cite{cattelan_modeling_2024}, naturally include these constraints in their formulation.

We encode the capacity constraint into our optimization model by using the recently proposed \textit{unbalanced penalization} strategy \cite{montanez-barrera_unbalanced_2024}. This method encodes inequality constraints by employing parabolic penalty functions that feature both linear and quadratic components. Importantly, this technique allows us to account for capacity constraints without the need for extra slack variables and has also demonstrated better performance compared to using slack variables in recent quantum annealing experiments \cite{montanez-barrera_improving_2023}. To encode the capacity constraint in \eqref{eq-capacity_constraint} using unbalanced penalization, we add the parabolic penalty
\begin{equation}
    H_k= -\lambda_1\left(C_{\text{max}}-\sum_i P_{i,k}\right) + \lambda_2 \left(C_{\text{max}}- \sum_i P_{i,k}\right)^2.
    \label{eq-unbalanced_penalization_Hamiltonian}
\end{equation}
where $\lambda_1$ and $\lambda_2$ are independent penalty scalars that can be modified to adjust the severity of the penalty. As the penalty constraint must be imposed on every partition, the full penalty Hamiltonian becomes
\begin{equation}
    H_{\text{pen}} = \sum_{k=0}^{d-1}H_k
\end{equation}
In Appendix \ref{appendix-cons_encoding}, we offer a comprehensive discussion on encoding capacity constraints via unbalanced penalization, including guidance on the appropriate selection of penalty scalars.

\subsection{Adaptive ansatz construction}
Rather than operating with a fixed ansatz, we propose a strategy of choosing the Hermitian operators included in the ansatz adaptively at each optimization step. For each qudit, we begin by introducing an operator pool
\begin{equation}
    P_i=\{G^i_l\}_{l=0}^{N_P^i-1}
\end{equation}
which we define as a set of $N_P^i$ Hermitian operators that generate unitary single-qudit gates on qudit $\ket{\psi_i}$. Although not written out explicitly, this implies that the operators $G^i_l$ must act as the identity operator on the remaining qudit states $\ket{\psi_j}$ where $j\neq i$. Next, we select the operator from the pool that generates a unitary operator maximizing the amplitude of the time gradient of the cost Hamiltonian. That is, we choose the Hermitian operator $G_i[s]$ generating a unitary on qudit $i$ at optimization step $s$ that satisfies
\begin{equation}
    G_i[s] = \arg\max_{G_l^i\in P_i}\left|\braket{\Psi_s|[G_l^i, H]|\Psi_s}\right|.
    \label{eq-construct_ansatz}
\end{equation}
where $\ket{\Psi_s}$ denotes the quantum state at optimization step $s$. Selecting the Hermitian operator in this manner is intended to prevent the inclusion of operators in the ansatz that minimally impact the expectation value of the cost Hamiltonian throughout the time evolution process. The process described above is performed individually for each qudit, selecting one Hermitian generator per qudit, thereby forming the entire ansatz. While related approaches have been explored for adaptively selecting mixer Hamiltonians in QAOA \cite{zhu_adaptive_2022}, applying such ideas in the context of QITE is, to the best of our knowledge, novel.

\subsection{Avoiding linear system of equations}
\label{sec-avoid_linsys}
In the implementation of QITE operating on product states of qubits with the linear ansatz $A_{\text{lin}=\{Y_i\}}$, one does not need to solve the linear system \eqref{eq-linear_system_of_eqs} to find the expansion coefficients. This is the case as $\braket{\Psi|Y_i Y_j|\Psi} = \delta_{i,j}$, given that $\ket{\Psi}$ is in a product state in which the individual qubits have no relative phase between the $\ket{0}$ and $\ket{1}$ states. Here, we build upon this idea to avoid solving the linear system of equations \eqref{eq-linear_system_of_eqs} in the qudit-based implementation. If the qudit states in the system only have real-valued expansion coefficients in the computational basis $\mathcal{B}_d$, and the Hermitian operators in the ansatz $A=\{G_i\}$ have imaginary-valued matrix elements in the computational basis, it must be the case that $\braket{\Psi|G_i|\Psi} = 0$ for every $G_i\in A$ at all times during state evolution. Assuming that we are operating with a linear ansatz such that $\ket{\Psi}$ always remains in a product state, we acquire the identity $\braket{\Psi|G_iG_j|\Psi} = \delta_{i,j}\braket{\Psi|G_i^2|\Psi}$. Utilizing this result, we are able to simplify \eqref{eq-linear_system_of_eqs} to 
\begin{equation}
    a_i = -\frac{i}{2}\frac{\braket{[G_i, H]}}{\braket{G_i^2}}
    \label{eq-avoid_linsys}
\end{equation}
thus allowing us to circumvent the need to solve a linear system of equations in the qudit-based implementation.

When solving the Min-$d$-Cut problem, we utilize the same operator pool for all qudits. In particular, we use an operator pool consisting of $d$ operators
\begin{equation}
    P = \left\{i\sum_{j=0, j\neq l}^{d-1}|j\rangle\langle l| +\text{h.c.}\right\}_{l=0}^{d-1}.
    \label{eq-pool_definition}
\end{equation}
Intuitively, we can think of the $l$:th operator in the pool in \eqref{eq-pool_definition} as coupling the $l$:th qudit computational basis state to the remaining computational basis states with equal coupling strength. Crucially, the operators in this pool all have imaginary matrix elements in the computational basis, thus implying that the expansion coefficients of the ansatz can be calculated using the simplified formula \eqref{eq-avoid_linsys}. We provide a concrete example of the operator pool in the case of qutrits, that is, $d=3$ qudit systems in Appendix \ref{appendix-qutrit_operators}.

\subsection{State propagation and relaxed rounding}
Once the expansion coefficients in \eqref{eq-avoid_linsys} have been determined, the state is updated as
\begin{equation}
    \ket{\Psi_{s+1}} =\bigotimes_{i=0}^{N-1}\exp(-ia_i[s] \Delta\tau G_i )\ket{\Psi_s}.
    \label{eq-state_update}
\end{equation}
with the initial state $\ket{\Psi_0}$ being given in \eqref{eq-initial_state_qudit}. In \eqref{eq-state_update}, the coefficients $a_i[s]$ are the result of calculating \eqref{eq-avoid_linsys} with all expectation values taken with respect to the current state $\ket{\Psi_s}$. The state update in \eqref{eq-state_update} effectively generates the quantum state $\ket{\Psi_{s+1}}$ by approximating the imaginary time evolution of the state $\ket{\Psi_s}$ over an imaginary time interval $\Delta\tau$. In our implementation, we propagate the quantum state using an imaginary time step of $\Delta\tau = 5\cdot10^{-3}$, as this appears to provide a good balance between stability and efficient convergence. The reliance on the first-order approximation in \eqref{eq-distance} for the calculation of the coefficients implies that a larger time step inevitably results in a greater error in the approximation.

During optimization, the quantum state $\ket{\Psi_s}$ is likely to be in a superposition of states corresponding to different classical integer strings. Since our primary aim is to identify a classical integer string that optimizes the objective in \eqref{eq-min_d_cut_cost_integer}, a single classical state must be derived from the quantum state. We acquire such a classical state by rounding each qudit state to the computational basis state for which it assumes the largest probability amplitude. That is, qudit $i$, represented by the state in \eqref{eq-qudit_state}, is assigned the integer $k_i$ given by
\begin{equation}
    k_i = \arg\max_{k}(\{|c_{i,k}|^2\}).
\end{equation}
\subsection{Complexity analysis and pseudocode} 
The computational bottleneck of the optimization protocol, as expected, is the process of determining the accurate expansion coefficients of the ansatz. In the calculation of any given expansion coefficient $a_i$ using \eqref{eq-avoid_linsys} we need only consider the terms in the Hamiltonian $H$ corresponding to edges connecting to vertex $i$, as only these terms can give a non-zero contribution in the commutator $[G_i, H]$. Consequently, each edge $(i,j)$ can only provide a non-zero contribution twice: once in the calculation of $[G_i, H]$ and once more during the calculation of $[G_j, H]$. During the evaluation of each such commutator, the $d$ different products of projectors in Max-$d$-Cut cost Hamiltonian in \eqref{eq-max_d_cut_cost_hamiltonian} must be considered. Therefore, the time complexity of the algorithm is $O(|E|d)$, where we let $|E|$ denote the edge count. In the worst-case scenario, the number of edges may increase quadratically relative to the number of vertices $N$, leading to a worst-case time complexity of $O(N^2d)$. 

We summarize the full qudit-based optimization protocol for $N_S$ optimization steps in Algorithm \ref{alg-wyd_qudit_alg}. In practice, the number of optimization steps $N_S$ does not require extensive fine-tuning and can simply be set to a sufficiently large upper limit. The iterative updates tend to produce long plateaus in which the objective changes only marginally, followed by occasional sudden decreases. A practical strategy is therefore to run the algorithm until the cost has remained stable over an extended sequence of iterations, at which point further increases in $N_S$ no longer improve the solution. We find that setting $N_S$ large enough to allow for such stabilization is typically sufficient.

\begin{algorithm}
\caption{Qudit-based QITE}
\label{alg-wyd_qudit_alg}
\begin{algorithmic}

\State $P_i \gets \{G_p^i\}_{p=0}^{N_P^i-1}$ 
\Comment{Define operator pools}
\vspace{0.4em}

\State $\ket{\Psi_0}\gets \ket{0}\bigotimes_{i=1}^{N-1} \frac{1}{\sqrt{d}}\sum_{k=0}^{d-1}\ket{k}$ 
\Comment{State initialization}
\vspace{0.4em}

\For{$s \in \{0,1,..., N_S-1\}$}
\vspace{0.4em}
    \State $\{G_i[s]\} \gets \left\{\arg\max_{P_i}  \left|i\braket{\Psi_s|[G_p^i, H]|\Psi_s}\right|\right\}_{i=0}^{N-1}$ %\Comment{Define ansatz}
    \vspace{0.4em}
    
    \State $a_i[s] \gets -\frac{i}{2}\braket{\Psi_s|[G_i[s], H]|\Psi_s}\braket{\Psi_s|G_i[s]^2|\Psi_s}^{-1}$% \Comment{Calculate expansion coefficients}
    %\State $G[s] \gets \sum_{i=0}^{N-1}a_i[s] G_i[s]$ \Comment{Define generator}
    \vspace{0.4em}
    
    \State $\ket{\Psi_{s+1}} \gets \bigotimes_{i=0}^{N-1}\exp(-ia_i[s] \Delta\tau G_i )\ket{\Psi_s}$ %\Comment{Evolve }
    \vspace{0.4em}
    
    \State $k_i \gets \arg\max_{k}\left( \{|c_{i,k}|^2\} \right)$ \Comment{Relaxed-rounding} 
    \vspace{0.4em}
\EndFor \\
\vspace{0.4em}
\Return $\{k_i\}_{i=0}^{N-1}$
\end{algorithmic}
\end{algorithm}

\section{Results and Discussion}
\subsection{Problem instances}
To evaluate the performance of the proposed qudit-based algorithm, we solve instances of the Min-$d$-Cut problem with capacity constraints on 10-regular graphs. Graphs are created by randomly generating vertex coordinates $(x,y)$, where both $x$ and $y$ are sampled uniformly at random from the range $[-1, 1]$. Next, each vertex is assigned an edge to its ten nearest neighbors, with edge weights assigned the rounded reciprocal distance
\begin{equation}
    W_{i,j} = \text{round}\left(\left(\sqrt{(x_i-x_j)^2+(y_i-y_j)^2}\right)^{-1}\right).
\end{equation}
Using this model, we expect clusters of vertices in close proximity to be assigned to the same partition, which also makes it straightforward to visually verify whether a solution is reasonable. Although the specific graph model may influence performance, exploring alternative graph families lies beyond the scope of this work. We solve the Min-$d$-Cut problem with $d=3,5,7$ on graphs with different vertex counts $N=50, 100, 150$. For each problem configuration, we consider 50 randomly generated graphs and benchmark against the Gurobi optimizer operating with a binary QUBO formulation. We choose to encode the problem in Gurobi using a QUBO formulation rather than as an integer program, as the presence of Kronecker deltas in \eqref{eq-min_d_cut_cost_integer} makes it impossible to express the cost as a quadratic mixed integer program of the form
\begin{equation}
    \min_{\mathbf{x}} \mathbf{x}^TQ\mathbf{x} + \mathbf{q}^T\mathbf{x}
\end{equation}
where $\mathbf{x}\in \mathbb{Z}^N$, $\mathbf{q}\in\mathbb{R}^{N}$, and $Q\in \mathbb{R}^{N\times N}$. For details on the binary Gurobi implementation, we refer to Appendix \ref{appendix-gurobi_encoding}.

As a metric for solution quality, we consider the approximation ratio (AR), calculated as the ratio of the cost provided by qudit-based QITE and Gurobi 
\begin{equation}
    {\text{AR}} = \frac{C_{\text{QITE}}}{C_{\text{Gurobi}}}.
\end{equation}
As we are solving a minimization problem, improved solutions provided by QITE are characterized by a lower AR. For the capacity constraint, we set the maximum number of vertices that can be assigned to any partition to $C_{\text{max}} = \frac{2N}{d}$. The capacity constraints are encoded using \textit{unbalanced penalization} \cite{montanez-barrera_unbalanced_2024}, as described in Appendix \ref{appendix-cons_encoding}.

\subsection{Numerical results}
\begin{figure}[ht]
  \centering
  \includegraphics[width=\columnwidth]{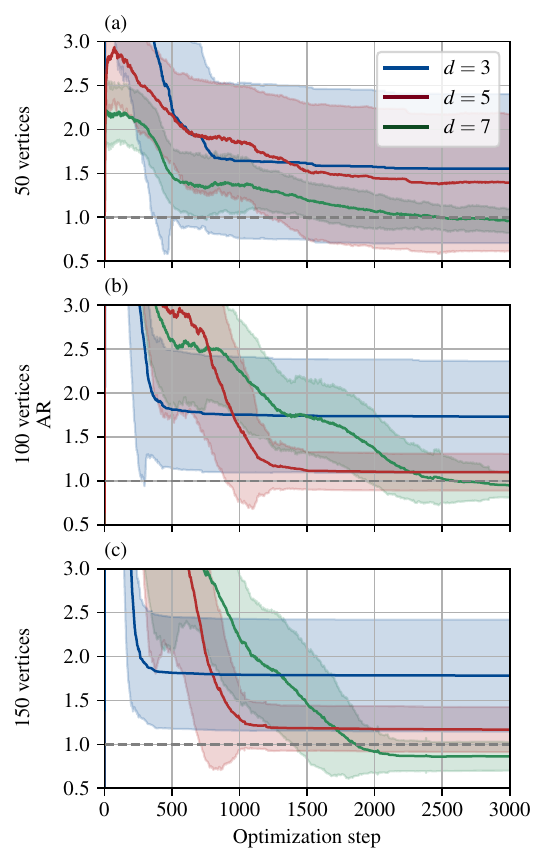}
  \caption{Mean approximation ratio (AR) as a function of the optimization step. 
The AR is calculated as the relative performance compared to Gurobi with 
capacity constraints encoded using unbalanced penalization. The color bands 
indicate the mean $\pm$ one standard deviation. Results are depicted separately 
for \textbf{(a)} 50, \textbf{(b)} 100, and \textbf{(c)} 150 vertices. 
The numerical results display particularly good performance for $d=7$ partitions, 
achieving an mean AR $< 1$ for all considered vertex counts.}
  \label{fig-AR}
\end{figure}
The numerical results of solving the Min-$d$-Cut problem are demonstrated in Fig. \ref{fig-AR}, with final approximation ratios summarized in Table \ref{tab-res_inequality_cons_AR}. The uncertainty intervals reported throughout represent one standard deviation across the 50 problem instances. Based on the results, we can observe particularly good performance of the qudit-based QITE implementation when solving the Min-$d$-Cut problem with $d=7$ partitions. Notably, we acquire an AR $< 1$ on average in the comparison against Gurobi, irrespective of the vertex count $N$ when $d=7$. The best results are achieved when solving with $N=150$ vertices and $d=7$ partitions, for which QITE provides an average AR of $0.857\pm0.147$. When considering $d=5$ partitions, the results are still promising, with an AR $1.099\pm0.209$ for $N=100$ vertices and AR $1.165 \pm 0.255$ for $N=150$ vertices. When considering only $d=3$ partitions, we can observe an increase in the AR, with the solution quality of QITE systematically falling short of Gurobi. Depending on the vertex count, the AR falls within the range $1.550\pm0.848$ to $1.783\pm0.637$, thus resulting in a significant performance drop compared to tested problem configurations with larger values of $d$.
\begin{figure}[ht]
  \centering
  \includegraphics[width=\columnwidth]{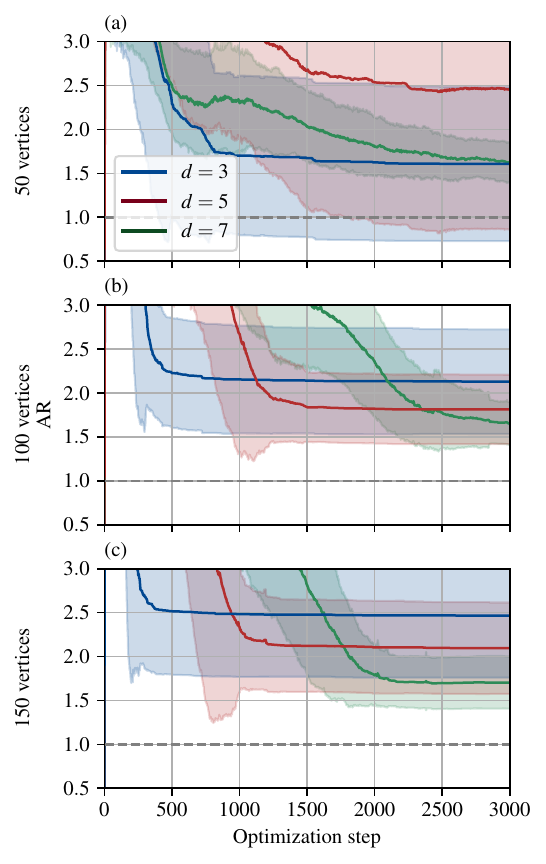}
  \caption{AR provided by QITE compared to Gurobi with capacity constraints encoded as hard constraints, for \textbf{(a)} 50, \textbf{(b)} 100, and \textbf{(c)} 150 vertices. QITE never acquires an AR superior to Gurobi on average when capacity constraints are encoded in Gurobi as hard constraints.}
  \label{fig-Gurobi_hard_AR}
\end{figure}

We speculate that a possible explanation for the enhanced performance at larger $d$ values might be attributed to the differences in the number of decision variables between the binary and integer-based encodings. The binary problem formulation requires $Nd$ decision variables, while the integer-based encoding only needs $N$ decision variables. As $d$ increases, the difference in the number of decision variables expands, thus increasing the significance of the integer-based encoding used by QITE. In addition to the lower AR achieved by QITE for larger values of $d$, we also observe a significant reduction in the standard deviation of the AR across problem instances (see Table \ref{tab-res_inequality_cons_AR}). We expect these trends to persist for moderately larger values of $d$, although not without limitations. Increasing $d$ further introduces two main challenges. First, it requires operating with a larger operator pool, which increases the computational cost of each time step. Second, enforcing the capacity constraint becomes more complicated, as additional penalty terms must be included in the Hamiltonian when the number of partitions grows. These effects suggest that while moderate increases in $d$ may continue to benefit QITE, very large $d$ values are expected to increase computational overhead and make constraint satisfaction more challenging.

\begin{table}[ht]
\renewcommand{\arraystretch}{1.4} % Increase row height
\centering
\caption{Mean approximation ratio ($\pm$ STD) by qudit-based QITE compared to Gurobi.}
\label{tab-res_inequality_cons_AR}
\begin{tabular}{c|ccc}
\multicolumn{4}{c}{\textbf{(a) Compared to Gurobi with penalty-encoded constraints.}} \\
\midrule
\textbf{Vertices} & $d=3$ & $d=5$ & $d=7$ \\
\midrule
50  & 1.550 ± 0.848 & 1.342 ± 0.753 & 0.922 ± 0.096 \\
100 & 1.727 ± 0.624 & 1.099 ± 0.209 & 0.950 ± 0.135 \\
150 & 1.783 ± 0.637 & 1.165 ± 0.255 & 0.857 ± 0.147 \\
\addlinespace[4pt]
\midrule
\multicolumn{4}{c}{\textbf{(b) Compared to Gurobi with hard-encoded constraints.}} \\
\midrule
\textbf{Vertices} & $d=3$ & $d=5$ & $d=7$ \\
\midrule
50  & 1.604 ± 0.879 & 2.383 ± 1.596 & 1.562 ± 0.168 \\
100 & 2.125 ± 0.589 & 1.814 ± 0.395 & 1.659 ± 0.253 \\
150 & 2.466 ± 0.707 & 2.088 ± 0.512 & 1.693 ± 0.290 \\
\bottomrule
\end{tabular}
\end{table}

In addition to benchmarking QITE against Gurobi with capacity constraints encoded via unbalanced penalization, we also compare its performance to Gurobi with hard capacity constraints. In this setting, as shown in Fig. \ref{fig-Gurobi_hard_AR} the relative performance of QITE decreases, with the average AR exceeding 1 across all considered problem configurations. However, similar to the penalized case, we observe that the AR decreases as the number of partitions increases. Most importantly, these results emphasize the critical role of efficient constraint encoding. Future progress in quantum and quantum-inspired optimization will therefore depend not only on advances in optimization strategies but also on the development of more effective methods for encoding constraints.

\begin{figure}
    \centering
    \includegraphics[width=\columnwidth]{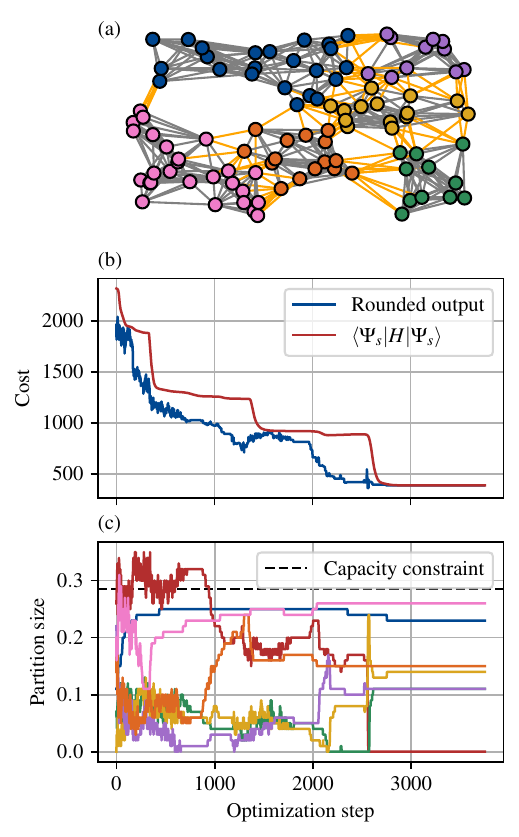}
    \caption{\textbf{(a)} Example output provided by qudit-based QITE on a graph with $N=100$ vertices, demonstrating clear clustering. \textbf{(b)} The cost of the rounded output and expectation value of the cost Hamiltonian as function of the number of optimization steps. Interestingly, the cost appears to reduce in a step-wise fashion. \textbf{(c)} Evolution of partition sizes throughout optimization, with the capacity constraint indicated by the horizontal dashed line. The colors of the partitions correspond to the colors of vertices in (a).}
    \label{fig-example_instance}
\end{figure}

Fig. \ref{fig-example_instance} demonstrates the optimization on one problem instance with $N=100$ vertices and $d=7$. We observe that the optimization process reduces the expectation value of the cost Hamiltonian in a step-wise manner. In particular, the trajectory is characterized by extended plateaus, where the objective makes little or no progress for many iterations, followed by sudden decreases. While the underlying cause of this behavior is not yet well understood, it appears to be a recurring feature of the optimization dynamics in our setting. We hypothesize that the observed behavior might be due to the gradual evolution of the state, suggesting that the same Hermitian operators might be selected from the operator pool over extended optimization sequences. Confirming if this is indeed the case would require an examination of the correlations between the stepwise energy reduction and the selection of operators from the pool, which we defer to future studies. Fig. \ref{fig-example_instance} (b) also allows us to compare the expectation value of the cost Hamiltonian to the cost of the classical integer string acquired by rounding the quantum state. Indeed, we see that the cost related to the rounded output consistently remains below the expectation value of the quantum state, verifying the effectiveness of the rounding method.

Finally, we briefly comment on the evolution of the sizes of the partitions during optimization. Fig. \ref{fig-example_instance} (c) depicts the sizes of the partitions throughout optimization as a fraction of the total number of vertices $N$. The sizes of the partitions seem to vary quite actively during the optimization process and then quickly stabilize to their final values as the expectation value of the cost undergoes its last step-like decrease.

\subsection{Adherence to capacity constraints}
\begin{figure*}
    \centering
    \includegraphics[width=\textwidth]{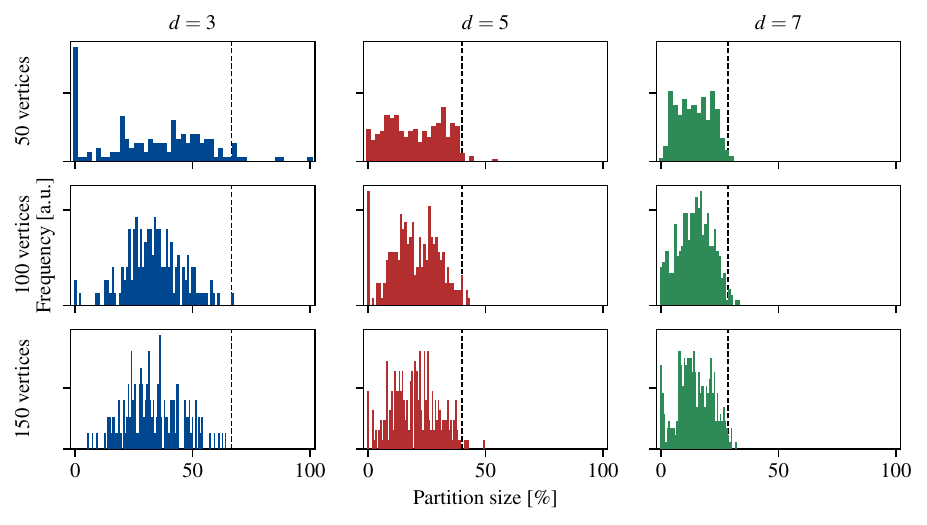}
    \caption{The distributions demonstrate the distribution of partition sizes, as a percentage of the total number of vertices in the graph and a resolution of one vertex. The qudit-based implementation of QITE systematically adheres to the capacity constraint of the Min-$d$-Cut problem, regardless the number of vertices and value of $d$. The maximum partition capacity $C_{\text{max}} = \frac{2N}{d}$ is indicated by the vertical dashed lines. Although the constraint is satisfied in almost all problem instances, many partitions are generated populated only by a small number of vertices. This issue appears slightly less severe for Gurobi, for which distributions are shown in the insets.}
    \label{fig-partition_size_distribution}
\end{figure*}

Next, we consider to what extent the qudit-based QITE implementation is able to adhere to the capacity constraints. Fig. \ref{fig-partition_size_distribution} demonstrates the distribution of the number of vertices assigned to partitions, presented separately for different values of $d$ and vertex counts $N$. For each configuration, the data represents the distribution of partition sizes accumulated over all 50 considered problem instances. If the optimizer successfully adheres to the encoded capacity constraints, the resulting distribution of partition sizes should remain bounded by the maximum permissible value $C_{\text{max}}$, ideally never exceeding this limit. As seen in Fig. \ref{fig-partition_size_distribution}, the optimizer only rarely produces partitions that violate the capacity constraint, and when violations occur, they exceed the limit by only a small margin. Crucially, the algorithm appears to satisfy the capacity constraint irrespective of the problem structure concerning the vertex count $N$ or the value of $d$, indicating robustness against the problem structure. On the other hand, the findings also seem to suggest that a significant number of partitions are populated by only a few vertices, which is contrary to what would intuitively be considered an optimal solution to the Min-$d$-Cut problem.

\begin{figure}
    \centering
    \includegraphics[width=\columnwidth]{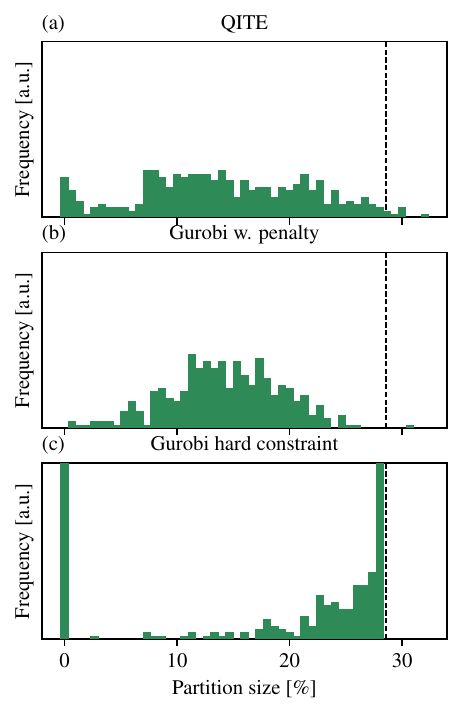}
    \caption{Distribution of partition sizes for $N=150$ vertices and $d=7$ partitions obtained with \textbf{(a)} QITE, \textbf{(b)} Gurobi with capacity constraints enforced via penalization, and \textbf{(c)} Gurobi with hard capacity constraints.
    QITE and Gurobi with penalty-based constraints yield qualitatively similar distributions, with no strong concentration near the capacity limit $C_{\text{map}}$ and few empty partitions. In contrast, Gurobi with hard constraints produces a distinct distribution, characterized by a peak near $C_{\text{max}}$ and a significant fraction of empty partitions. This structural difference in the distributions helps explain the superior performance of Gurobi when using hard capacity constraints.}
    \label{fig-distribution_comparison}
\end{figure}

It is also instructive to examine how the distribution produced by QITE compares to that obtained with Gurobi. In Fig. \ref{fig-distribution_comparison}, we show the distribution of partition sizes for $N=150$ vertices and $d=7$ partitions. Panels (b) and (c) display the distributions obtained with Gurobi when capacity constraints are enforced through penalization and through hard constraints, respectively.

When comparing QITE to Gurobi with penalty-based constraints, the resulting distributions appear qualitatively similar: in both cases, there is neither a strong concentration of partitions near the capacity limit $C_{\text{map}}$ nor a large number of empty partitions. In contrast, Gurobi with hard capacity constraints yields a markedly different distribution, characterized by a pronounced peak near $C_{\text{max}}$ and a substantial number of unpopulated partitions. This difference in distribution likely plays a central role in the superior performance of Gurobi when hard constraints are imposed, as observed earlier.

\section{Conclusion}
Qudit-based formulations present promising opportunities for quantum-inspired optimization. Unlike quantum hardware, classical algorithms can utilize higher-dimensional qudit encodings for more efficient problem representation. In this work, we propose an optimization protocol for combinatorial problems with integer-valued decision variables, generalizing previous methods for approximating imaginary time evolution on qubits to qudits. Each classical variable is encoded in a qudit, and the state evolves through single-qudit unitaries. We also introduce a gradient-based approach for adaptively selecting the ansatz at each optimization step, eliminating the need to solve linear systems for state updates.

To benchmark our proposed solution, we solved several constrained instances of Min-$d$-Cut problem on randomly generated graphs for different values of $d$ and varying vertex counts. When using the same unbalanced parabolic penalization to encode constraints in both QITE and Gurobi, QITE provided superior solutions on average in problem instances for which $d=7$. On the other hand, the relative performance compared to Gurobi dropped significantly when considering problem configurations with smaller values of $d$, indicating that qudit-based QITE is sensitive to the structure of the underlying problem. Moreover, QITE, when using a penalized constraint formulation, underperformed relative to Gurobi with native encoding of hard capacity constraints across all problem configurations, underscoring the need for efficient constraint encoding in quantum optimization.

Future work could evaluate qudit-based QITE across a wider range of optimization problems to identify those that are challenging or well-suited for our approach. Additionally, exploring alternative pools of Hermitian operators in adaptive ansatz construction could offer valuable insights. Specifically, tuning the coupling strengths between computational basis states, which were set to unity in the current formulation, may enhance the flexibility of quantum state evolution.

\section{Acknowledgments}
This manuscript is based on work carried out as part of the author’s master’s thesis \cite{marton2025quantum}.

%\printbibliography
\bibliography{references}

\newpage
\onecolumn
\appendix
\subsection{Example of qutrit operator pool}
\label{appendix-qutrit_operators}
The specific operator pool utilized in this work is defined in \eqref{eq-pool_definition}. Here, we provide a concrete example of the operator pools for the specific case of qutrits, that is, $d=3$ qudit states. Following the definition of \eqref{eq-pool_definition}, the operator pool for this case would consist of the following operators

\begin{equation}
\begin{cases}
    G_0 = i(\ketbra{1}{0} + \ketbra{2}{0}) + \text{h.c.} \\
    G_1 = i(\ketbra{0}{1} + \ketbra{2}{1}) + \text{h.c.} \\
    G_2 = i(\ketbra{0}{2} + \ketbra{1}{2}) + \text{h.c.}.
\end{cases}
\label{eq-qutrit_pool}
\end{equation}
Alternatively, we can consider the operators in \eqref{eq-qutrit_pool} in their matrix representation in the computational basis $\mathcal{B}_3$
\begin{equation}
    \begin{cases}
        G_0 = i\begin{bmatrix} 
        0 & -1 & -1\\
        1 & 0 & 0\\
        1 & 0 & 0
\end{bmatrix}_{\mathcal{B}_3} \\
        G_1 = i\begin{bmatrix} 
        0 & 1 & 0\\
        -1 & 0 & -1\\
        0 & 1 & 0
\end{bmatrix}_{\mathcal{B}_3} \\
        G_2 = i\begin{bmatrix} 
        0 & 0 & 1\\
        0 & 0 & 1\\
        -1 & -1 & 0
\end{bmatrix}_{\mathcal{B}_3}.
    \end{cases}
\end{equation}

\subsection{Binary problem encoding in Gurobi}
\label{appendix-gurobi_encoding}
We encode the Min-$d$-Cut problem in Gurobi using a binary formulation, as the occurrence of Kronecker deltas in the cost of the integer formulation in \eqref{eq-min_d_cut_cost_integer} implies that it cannot be formulated as a mixed-integer program. In the binary formulation, we introduce the decision variables $\{x_{i,k}\}$, where $x_{i,k}\in\{0,1\}$. We interpret $x_{i,k}=1$ as vertex $i$ having been assigned to partition $k$, and vice versa. As each vertex is effectively represented using $d$ binary variables, a total of $Nd$ decision variables are required to encode the Min-$d$-Cut on a graph with $N$ vertices. 

The cost of the Min-$d$-Cut problem is formulated as
\begin{equation}
    C(\textbf{x}) = \sum_{\langle i,j \rangle}W_{i,j}\left(1-\sum_k x_{i,k}x_{j,k}\right)
\end{equation}
where every vertex must also satisfy the single association constraint 
\begin{equation}
    \sum_k x_{i,k}=1.
\end{equation}
The single association constraints are encoded as hard constraints within the Gurobi model. We argue that this allows for a fair comparison against qudit-based QITE, for which the single association constraint is natively encoded in the qudit-based encoding and is hence guaranteed to be satisfied. The capacity constraint is defined as
\begin{equation}
    \sum_{i}x_{i,k}\leq C_{\text{max}}.
    \label{eq-capacity_constraint_appendix}
\end{equation}
and must be satisfied for every partition $k$. Here, $C_{\text{max}}$ denotes the maximum number of vertices that can be assigned to any given partition. In order to provide a thorough assessment of QITE, we impose the capacity constraint in \eqref{eq-capacity_constraint_appendix} using both unbalanced penalization for every partition $k$ (see Appendix \ref{appendix-cons_encoding}) and by encoding them as a hard constraint in the Gurobi optimization model. When encoding the capacity constraints using unbalanced penalization, we ensure that the values of the penalty scalars are the same as for QITE in order to allow for a direct comparison.

\subsection{Capacity constraint encoding using unbalanced penalization}
\label{appendix-cons_encoding}
\begin{figure}
    \centering
    \includegraphics[width=252pt]{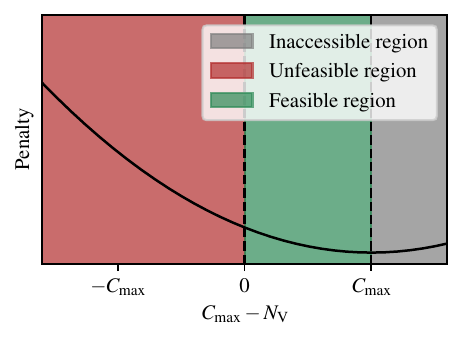}
    \caption{Unbalanced penalization encodes the capacity constraint using a parabolic penalty. Since the number of vertices $N_V$ assigned to any given partition is non-negative and the penalty scalars are chosen such that the minimum penalty is enforced when $N_V = 0$, the penalty is effectively given by a half-parabola.}
    \label{fig-unbalanced_penalization}
\end{figure}

The capacity constraint enforces that a maximum of $C_{\text{max}}$ vertices are allowed to be assigned to any partition $k$. In the integer encoding, this is formulated as
\begin{equation}
    \sum_i\delta_{x_i,k} = C_{\text{max}}.
    \label{eq-capacity_constraint_01}
\end{equation}
We can encode this constraint into the optimization model by utilizing unbalanced penalization \cite{montanez-barrera_unbalanced_2024}. In order to encode the constraint \eqref{eq-capacity_constraint_01} for partition $k$, we apply the parabolic penalization
\begin{equation}
    P_k^{\text{int}} = -\lambda_1\left(C_{\text{max}}-\sum_i \delta_{x_i,k}\right) + \lambda_2 \left(C_{\text{max}}- \sum_i \delta_{x_i,k}\right)^2.
    \label{eq-unbalanced_penalization_cost}
\end{equation}

Furthermore, as the constraint must be satisfied by all partitions, the full penalty $P^{\text{int}}$ becomes
\begin{equation}
    P^{\text{int}}=\sum_{k=0}^{d-1}P_k^{\text{int}}
    \label{eq-full_penalty}
\end{equation}
In \eqref{eq-unbalanced_penalization_cost}, $\lambda_1$ and $\lambda_2$ are both tunable penalty scalars that can be chosen independently. In our implementation, we fix the ratio as
\begin{equation}
    \frac{\lambda_1}{\lambda_2} = 2C_{\text{max}}
    \label{eq-lambda_ratio}
\end{equation}
since this ensures that the minimum of the parabolic penalty is enforced when a partition is left unpopulated. To encode the parabolic penalty \eqref{eq-unbalanced_penalization_cost} in QITE, the Kronecker deltas in \eqref{eq-unbalanced_penalization_cost} are replaced by the projectors $P_{i,k} = |k\rangle_i\langle k|_i$, thus giving the penalty Hamiltonian
\begin{equation}
    H_k^{\text{int}} = -\lambda_1\left(C_{\text{max}}-\sum_i P_{i,k}\right) + \lambda_2 \left(C_{\text{max}}- \sum_i P_{i,k}\right)^2.
    \label{eq-unbalanced_penalization_Hamiltonian_1}
\end{equation}
In the binary formulation of the Min-$d$-Cut problem used in Gurobi, the capacity constraint on partition $k$ is instead encoded by adding the penalty term
\begin{equation}
    P_k^{\text{bin}} = -\lambda_1\left(C_{\text{max}}-\sum_i x_{i,k}\right) + \lambda_2 \left(C_{\text{max}}- \sum_i x_{i,k}\right)^2.
    \label{eq-unbalanced_penalization_cost_binary}
\end{equation}
Empirically, we find that larger values of the penalty scalars are required when optimizing with a larger value of $d$. We set the penalty scalar corresponding to the linear term as $\lambda_1 = 5$ for $d=3$, $\lambda_1 = 20$ for $d=5$, and $\lambda_1 = 30$ for $d=7$. The penalty scalar $\lambda_2$ corresponding to the quadratic term is acquired using \eqref{eq-lambda_ratio}.

As the number of vertices $N_V$ assigned to any given partition is necessarily non-negative, this implies that unbalanced penalization effectively encodes the capacity constraint using a half-parabola, since half of the parabolic penalty will remain inaccessible. This is demonstrated in Fig. \ref{fig-unbalanced_penalization}.
\end{document}